\documentclass[twocolumn,pra]{revtex4}

\usepackage{graphicx}
\usepackage{dcolumn}
\usepackage{bm}
\usepackage{epsfig}
\usepackage{footnote}
\usepackage{times}

\begin{document}

\title{Noise-free high-efficiency photon-number-resolving detectors}

\author{Danna Rosenberg}
\author{Adriana E. Lita}%
\author{Aaron J. Miller}%
\author{Sae Woo Nam}%
\affiliation{National Institute of Standards and Technology,
$325$ Broadway, Boulder, Colorado 80305, USA}%

\date{Received 23 March 2005; revised 15 April 2005; published 17 June 2005}

\begin{abstract}

High-efficiency optical detectors that can determine the number of
photons in a pulse of monochromatic light have applications in a
variety of physics studies, including post-selection-based
entanglement protocols for linear optics quantum computing and
experiments that simultaneously close the detection and
communication loopholes of Bell's inequalities.  Here we report on
our demonstration of fiber-coupled, noise-free,
photon-number-resolving transition-edge sensors with $88~\%$
efficiency at $1550$~nm.  The efficiency of these sensors could be
made even higher at any wavelength in the visible and
near-infrared spectrum without resulting in a higher dark-count
rate or degraded photon-number resolution.
\end{abstract}


\maketitle

High-efficiency, photon-number-resolving detectors can transform
the field of quantum optics. One of many experiments they can
enable is linear optics quantum computing, which requires
postselection based on photon number \cite{KLM}. So far,
researchers have succeeded in implementing simple two-qubit gates
by using conventional detectors that cannot distinguish between
one and two photons \cite{Gasparoni04,branning04}, but moving
beyond two qubits requires high-efficiency, low dark-count rate,
photon-number-resolving detectors. Such detectors can be used to
herald multiphoton path-entangled states from the output of a
parametric down-conversion crystal \cite{Bouwprl05}, and these
entangled states could be used for applications ranging from
quantum cryptography \cite{Ekert91,BBM92} to lithography beyond
the diffraction limit \cite{Boto}\footnote{In order to be
practical, quantum lithography requires the development of a {\it
bright} source of entangled photons.}. Photon-number-resolving
detectors can also be used to verify the quality of single-photon
sources, a necessity for secure information transfer in some
quantum key distribution protocols. Furthermore, low-noise,
high-efficiency detectors that operate at telecommunication
wavelengths can significantly extend the length of a secure link
in fiber quantum key distribution where light must be transmitted
over large distances \cite{lut99}.

Conventional detectors that operate in the visible and
near-infrared, such as avalanche photodiodes and photomultiplier
tubes, may be single-photon sensitive, but they cannot reliably
determine the number of photons in a pulse of
light\cite{APD,PMT,IRAPD}. In principle, beam-splitters and
single-photon sensitive detectors can simulate a photon-number
resolving detector \cite{franson04}, but the probability of
correctly identifying an $N$-photon event drops exponentially with
$N$, even for detectors with $100~\%$ efficiency, because it is
impossible to control the path each photon takes at a
beamsplitter.  Novel technologies such as the visible light photon
counter \cite{Waks03} have some photon-number resolution ability,
but operating them at maximum detection efficiency introduces dark
counts at rates greater than 10 kHz, and these detectors are
sensitive in the visible spectrum only.

\begin{figure}
\includegraphics[width=2.75in]{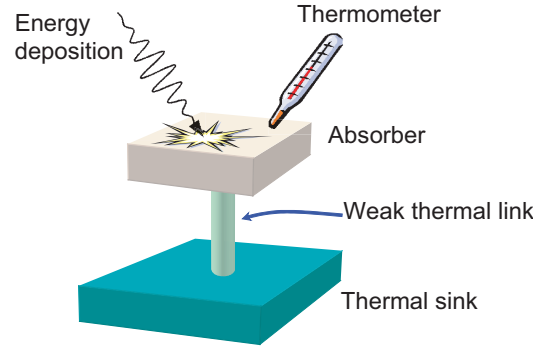}
\caption{Calorimeter operation. Energy is deposited in the
absorber and the thermometer reads out the change in temperature.
The absorber cools down slowly through the weak link to the
thermal heat sink. The tungsten transition-edge sensors discussed
in this paper are quantum calorimeters, with the tungsten
electrons acting as both the absorber and thermometer.  A voltage
bias keeps the tungsten electrons on the edge of the
superconducting-to-normal transition, and the steep dependence of
resistance on temperature provides precise thermometry.  The
anomalously low electron-phonon coupling in tungsten is the weak
thermal link.\label{fig:calor}}
\end{figure}

Superconducting transition-edge sensors (TESs) have photon-number
resolution with negligibly-low dark counts. The TESs discussed in
this paper are quantum calorimeters optimized for detection of
near-infrared and visible photons
\cite{Irwin95,Cabrera98,Miller03}.  The main components of a
calorimeter are the absorber, a thermometer, and a weak link to a
thermal heat sink, as shown in Fig. \ref{fig:calor}. When energy
impinges on the absorber, it heats up quickly and slowly cools
through the weak thermal link, and the temperature change is
measured by the thermometer. Detection of visible and
near-infrared light at the single-photon level places stringent
requirements on the heat capacity and thermometry.  For the TESs
described here, the electron subsystem in a thin film of tungsten
plays the parts of both the absorber and thermometer. The detector
is cooled below its superconducting transition temperature and a
voltage bias is applied to increase the electron temperature above
that of the substrate. At low temperatures, the electrons in
tungsten have anomalously low thermal coupling to the phonons,
providing the weak thermal link, and the rapid change in
resistance near the superconducting critical temperature results
in a very sensitive measure of temperature. The temperature change
due to energy deposition by a photon results in a change in
resistance, and the current change in the voltage-biased detector
is measured with a superconducting quantum-interference device
(SQUID) array \cite{Huber01}. The change in temperature (and thus
current) is proportional to the photon energy, so the sensor can
resolve the number of photons in a pulse of monochromatic light.

\begin{figure}
\includegraphics[width=2.75in]{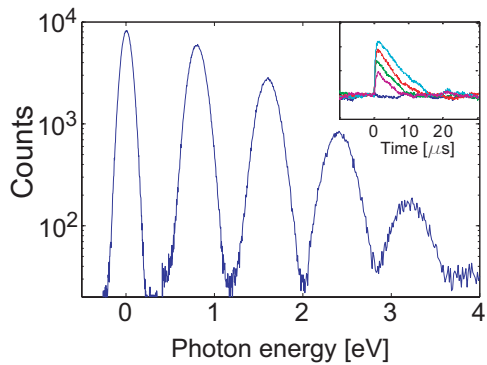}
\caption{Histogram of corrected pulse heights.  The sensor was
illuminated with a $1550~{\rm nm ~(0.8 eV)}$ gain-switched laser
diode pulsed at $50$~kHz with a pulse duration of $4$~ns. The
pulses from the detector were shaped with a $2~\mu$s semi-Gaussian
filter and the maximum value for each pulse was determined. (For
events in the zero-photon peak, the amplitude of the signal was
measured at the expected arrival time of the photon.) The
relationship between filtered pulse height and energy was
determined by measuring the mean of each histogram peak, and the
data were then corrected to linearize the response of the
detector. The full width at half maximum of the zero-, one-, two-,
three-, and four-photon peaks are $0.13$~eV, $0.20$~eV, $0.25$~eV,
$0.34$~eV, and $0.45$~eV, respectively. The inset shows typical
unfiltered pulses for zero-, one-, two-, three- and four-photon
events. \label{fig:hist}}
\end{figure}

The detection efficiency of a bare thin film tungsten sensor
$20~\rm{}nm$ thick is $15$ to $20~\%$ at visible and near-infrared
wavelengths and is limited by reflection from the front surface
and transmission through the film. However, every photon that is
absorbed by the tungsten leads to a change in temperature of the
electrons, so the detection probability can be increased by
embedding the tungsten detector in a stack of optical elements
that enhance the absorption of the light in the tungsten. The TESs
discussed in this letter measure $25~\mu{\rm m}$ by $25~\mu{\rm
m}$ and are approximately $20~\rm{}nm$ thick with superconducting
critical temperatures of ${\rm 110 \pm 5~mK}$. They are embedded
in structures that are designed to maximize absorption at
$1550~\rm{}nm$, a wavelength of particular interest for
telecommunications. Existing semiconductor-based detectors have
low ($10-20~\%$) efficiency and high (${\rm 10-20~kHz}$)
dark-count rates at this wavelength.  A detailed description of
the structures is presented elsewhere \cite{asc04}.  These sensors
have thermal decay times as short as $5~\mu s$ and provide
excellent discrimination between multi-photon events.  Figure
\ref{fig:hist} shows data from a TES embedded in an optical
structure designed to enhance the absorption of light at a
wavelength of $1550$~nm. The sensor was illuminated with a pulsed
source of $\rm{}1550~nm$ photons. The histogram displays the
distribution of pulse heights after the data were corrected for
the non-linearity in the temperature dependence of the resistance
in the superconducting transition. The full width at half maximum
(FWHM) of the zero-, one-, two-, three- and four-photon peaks are
$0.13$~eV, $0.20$~eV, $0.25$~eV, $0.34$~eV, and $0.45$~eV,
respectively. The increase in the FWHM of the peaks with
increasing energy is due to the device non-linearity mentioned
above. Measurements and simulations of the optical properties of
the various layers indicate that the total expected efficiency of
the sensor, neglecting system losses, is $92~\%$.

Photons were coupled to the detector through $9~\mu\rm{}m$ core
single-mode fiber with an anti-reflective coating for
$1550~\rm{}nm$. The fiber was held $50-75~\mu\rm{}m$ above the
detector and aligned at room temperature by backside through-chip
imaging. Focusing the light from the fiber was not necessary
because the spot size was small enough at this distance that
greater than $99~\%$ of the light was incident on the detector.
The housing holding the fiber was clamped in place and the
detector was then cooled to less than $100~\rm{}mK$ in an
adiabatic demagnetization refrigerator.  Because the applied
voltage bias keeps the electrons in the superconducting-to-normal
transition, the detector is not sensitive to slight fluctuations
in the cryostat temperature as long as the temperature is well
below the superconducting transition temperature of ${\rm 110 \pm
5~mK}$.

Coupling and alignment losses reduce the measured efficiency of
the detector from the expected $92~\%$. To minimize connection
losses, the fiber from the detector and the fiber going to room
temperature were fused together in the cold space of the cryostat.
The typical loss for a fiber fuse is approximately $0.5~\%$. We
measured the room temperature loss from outside the cryostat to
the sample space to be $2.3~\%$. Tests to determine the loss in a
loop of fiber that passes through the cold space of the
refrigerator indicated that the loss does not change when the
fiber is cooled. Thermal cycling of the fiber-coupled detector did
not change its efficiency, and we measured greater than $80~\%$
efficiency for several different detectors, indicating that our
alignment method is robust and that the fiber-to-detector
alignment does not degrade when cooled.

\begin{figure}
\includegraphics[width=2.75in]{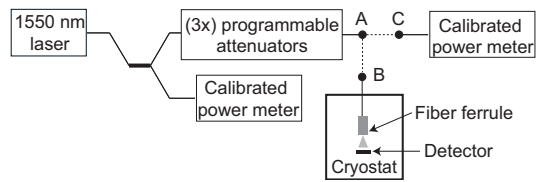}
\caption{Experimental apparatus for the detector efficiency
measurement. Light from the laser was split into two paths by a
fiber coupler. The top path contained three programmable
attenuators. A calibrated power meter at point A was used to
measure the actual attenuation value at every attenuation set
point used in the experiment. After the attenuators were
calibrated, the fiber exiting the last attenuator was cut at point
A and fused to the fiber (point B) leading to the detector. The
power in the bottom arm of the fiber coupler was continuously
monitored to ensure that the laser power did not drift.  All
measurements were performed within the linear range of the power
meter. \label{fig:calib}}
\end{figure}

Measuring the efficiency of the detectors is nontrivial due to the
low power levels involved and the introduction of loss through
fiber connectors.  The relatively slow pulse decay (several
microseconds) and the desire to avoid pulse pile-up requires the
use of subfemtowatt average optical power levels.  At present,
commercial power meters do not have the sensitivity required to
measure such low levels.  To circumvent this problem, we
calibrated a series of programmable optical attenuators using a
calibrated power meter well within its linear regime, as shown in
Fig. \ref{fig:calib}. The attenuator calibration and efficiency
measurements were performed using a laser with a center wavelength
of $1550~\rm{}nm$ and FWHM of $0.05~{\rm nm}$.

\begin{figure}
\includegraphics[width=2.75in]{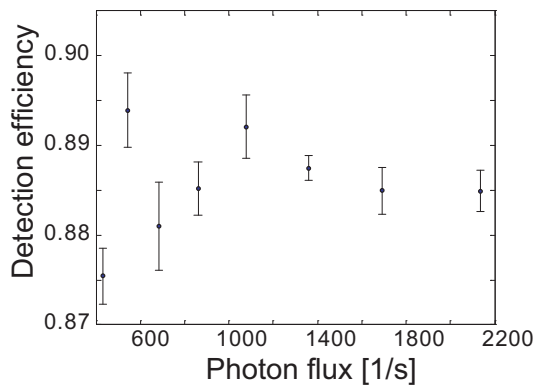}
\caption{Detection efficiency. The efficiency was measured at
several different average power levels with the laser in
continuous-wave operation. The error bars show the statistical
counting errors only. The slope of a weighted linear fit to the
data is zero within the error bars, indicating that there is no
dependence of efficiency on power level.  A weighted average of
the data yields an efficiency of $88.6~\%$ with a combined
uncertainty of $0.4~\%$. \label{fig:qe}}
\end{figure}

The efficiency measurements presented here were performed in
continuous-wave operation at several different power levels to
ensure that there was no dependence of measured efficiency on
power level, as shown in Fig. \ref{fig:qe}. Power levels were
adjusted by means of the calibrated programmable attenuators. The
number of pulses with pulse heights within $\pm 3\sigma$ ($\sigma
\approx 0.07~\rm{}eV)$ of the one-photon peak was recorded at each
power level for $100~\rm{}s$. The background rate in the same
energy range, which was approximately $400~\rm{}Hz$ and was due to
blackbody radiation from room temperature surfaces
\cite{tobepub}\footnote[1]{Note that the background counts were
not from detector noise; rather, they resulted from photons from
the low-energy tail of the blackbody distribution propagating
through the optical fiber.}, was measured periodically between
data sets and subtracted from the raw count rate. The procedure
outlined above provides a measure of the number of single-photon
events from the laser. At the higher power levels, counts were
present at energies greater than $3\sigma$ above the mean energy
of the one-photon peak. The spectral weight at these energies
results from pulse pile-up, the arrival of a second photon before
the system has recovered from a previous event. To correct for
this effect, we included the high-energy counts as two-photon
events. This correction is smallest ($1~\%$ of the total counts)
at the lowest power levels and remains below $2~\%$ at even the
highest power level.

Fig. \ref{fig:qe} shows the detection efficiency as a function of
power level, with error bars given by the uncertainties due to
Poisson statistics.  Not shown are the uncertainties resulting
from fiber bends at room temperature.  We have observed that small
bends in the fiber can easily lower the measured efficiency by up
to three per cent, and the scatter in efficiency measurements is
most likely due to slight shifts in the fiber position. All the
efficiency values presented are relative to the amount of light in
the fiber at point A in Fig. \ref{fig:calib}.

The measured system efficiency of $88.6 \pm 0.4 ~\%$ is consistent
with measurements and simulations of the optical elements and the
system losses. This detection efficiency exceeds the threshold of
$83~\%$ required to close the detection loophole in an experiment
testing Bell's inequalities.  This enables an experiment that
would simultaneously close both the detection and communication
loopholes, decisively refuting a local realism interpretation of
quantum mechanics.

Increasing the detection efficiency beyond $88.6~\%$ at
$1550~\rm{}nm$ is in principle simple and involves fabricating an
optical structure with more layers and finer control over layer
thickness. Similarly, it should be possible to produce near
unity-efficiency detectors at any wavelength in the ultraviolet to
near-infrared frequency range with this technique. Simulations
indicate the possibility of increasing the efficiency well above
$99~\%$ at any given wavelength in this spectrum, making these
detectors an extremely valuable tool for quantum optics and
quantum information processing.

\begin{acknowledgments}
The authors thank ARDA for financial support, Alan
 Migdall, Richard Mirin, John Martinis, Alexander Sergienko and
Erich Grossman for valuable technical discussions, and Marty Gould
of Zen Machining.  D. R. is supported by the DCI postdoctoral
program.
\end{acknowledgments}

\end{document}